\documentclass[nato,numreferences]{crckbked}
\usepackage{epsfig}

\begin{document}

\begin{article}
\begin{opening}

\title{An effective model of the spacetime foam}

\author{Vladimir \surname{Dzhunushaliev}
\thanks{\email{dzhun@hotmail.kg}}}

\institute{Kyrgyz - Russian Slavic University, Bishkek, Kyrgyzstan}

\runningauthor{Vladimir Dzhunushaliev}
\runningtitle{An effective model of the spacetime foam}

\begin{abstract}
An approximate model of the spacetime foam is offered in which
each quantum handle (wormhole) is a 5D wormhole-like solution.
A spinor
field is introduced for an effective description of this foam.
The topological handles of the spacetime foam can be
attached either to one space or connect two different spaces.
In the first case we have a wormhole with the quantum throat and such
object can demonstrate a model of preventing the formation
the naked singularity with relation $e > m$. In the second
case the spacetime foam looks as a dielectric with quantum
handles as dipoles. It is supposed that supergravity
theories with a nonminimal interaction between spinor and
electromagnetic fields can be considered as an effective model
approximately describing the spacetime foam.
\end{abstract}

\end{opening}

\section{Introduction}

\begin{figure}
\centerline{ \framebox{
\epsfig{figure=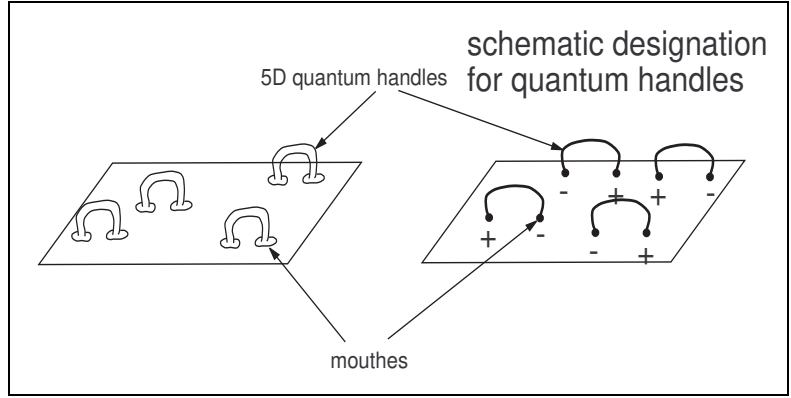,height=5cm,width=10cm}}}
\label{dzh-fig1}
\caption{At the left side of the figure is presented
a hypothesized spacetime foam. If we neglect of the cross
section of handle then (at the right) hand we have a schematic
designation for the spacetime foam.}
\end{figure}

The notion of a spacetime foam was introduced by Wheeler
\cite{wheel0,wheel1} for the
description of the possible complex structure of the spacetime on
the Planck scale ($L_{Pl} \approx 10^{-33}cm$). This hypothesized
spacetime foam is a set of quantum wormholes (WH) (handles) appearing
in the spacetime on the Planck scale level (see Fig.\ref{dzh-fig1}).
For the macroscopic observer these quantum fluctuations are
smoothed and we have an ordinary smooth manifold with the metric
submitting to Einstein equations. The exact mathematical
description of this phenomenon is very difficult and even though
there is a doubt: does the Feynman path integral in the gravity
contain a topology change of the spacetime ? This question spring
up because (according to the Morse theory) the singular points
must arise by the topology change. In such points the time arrow
is undefined that leads in difficulties at definition of the
Lorentzian metric, curvature tensor and so on.
The main goal of this paper is to submit
\textit{\textbf{an effective model of the spacetime foam.}}

\section{Model of a single quantum wormhole}

\begin{figure}
\centerline{ \framebox{
\epsfig{figure=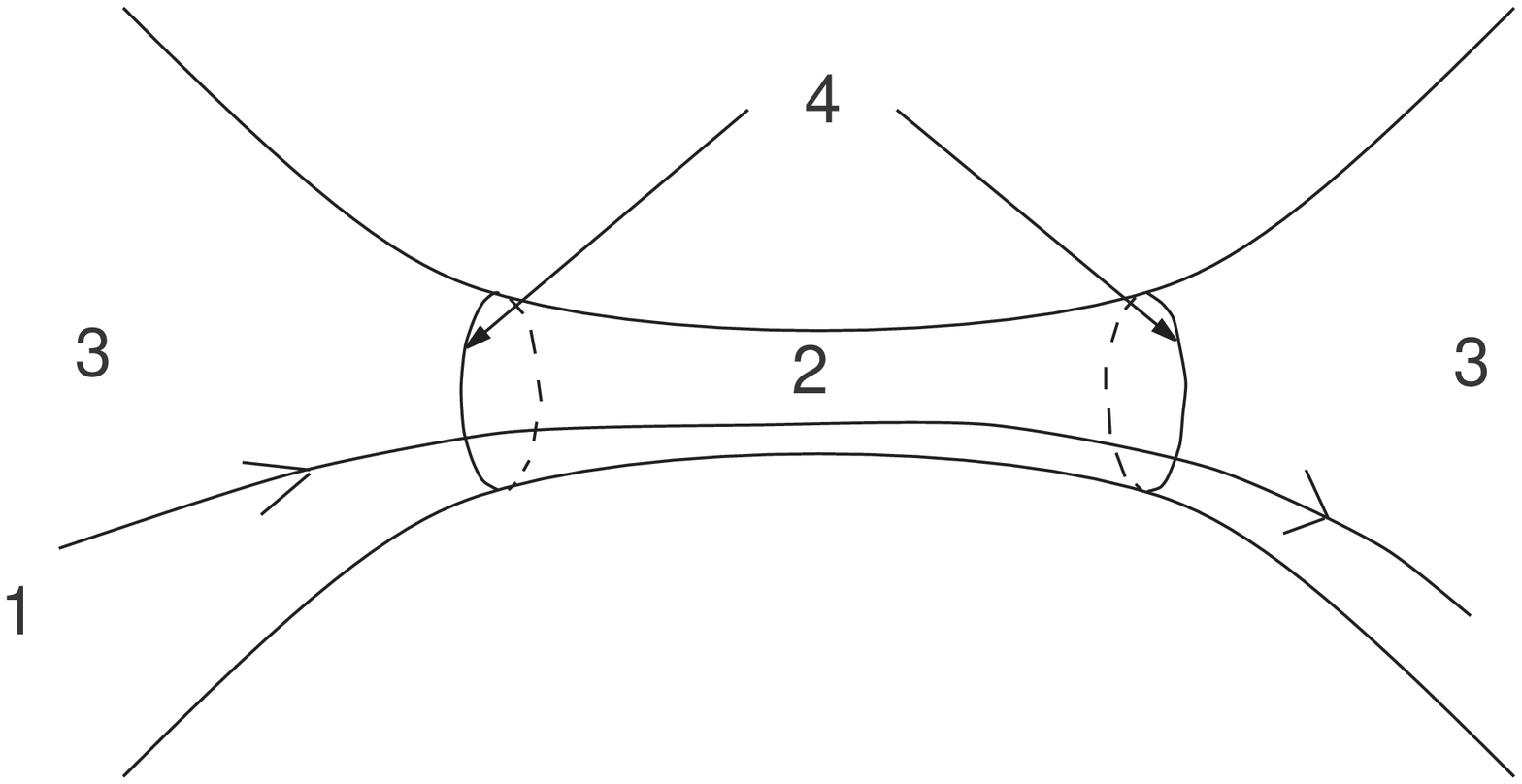,height=5cm,width=10cm}}}
\label{dzh-fig2}
\caption{
Here whole spacetime is 5D but in the external spacetime
(3) $G_{55}$
is nonvariable and we have Kaluza-Klein theory in its initial
interpretation as 4D gravity + electromagnetism. In the throat
(2) $G_{55}$ component of the 5D metric is equivalent to 4D
gravity+electromagnetism+scalar field. Near the event horizon
(4) the metric is the Reissner-Nordstrom metric
and the throat is a solution of the 5D Kaluza-Klein theory.
We should join these metrics on the event horizons.
(1) is the force line of the electric field.
}
\end{figure}
At first we present a model of a single
handle in the spacetime foam, see Fig(\ref{dzh-fig2}).
The 5D metric \cite{chodos,clement,dzhun97} for the throat is
\begin{eqnarray}
ds^{2} & = & \eta_{AB}\omega ^A\omega ^B =
\nonumber \\
&& -
\frac{r_0^2}{\Delta (r)}(d\chi  - \omega (r)dt)^2 + \Delta
(r)dt^{2} - dr^{2} -
\nonumber \\
&&
a(r) \left ( d\theta ^2 + \sin\theta ^2
d\varphi ^2 \right ),
\label{dzh-eq1}\\
a & = & r^{2}_{0} + r^{2},
\quad \Delta = \pm \frac{2r_0}{q}\frac{r^2 + r_0^2} {r^2 - r_0^2}
, \quad
\nonumber \\
&&
\omega = \pm \frac{4r_0^2}{q}\frac{r} {r^2 - r_0^2} .
\label{dzh-eq2}
\end{eqnarray}
where $\chi $ is the 5$^{th}$ extra coordinate;
$\eta_{AB} = (\pm,-,-,-,\mp)$, $A,B = 0,1,2,3,5$;
$r,\theta ,\varphi $ are the $3D$  polar coordinates;
$r_0 > 0$ and $q$ are some constants.
We can see that there are two closed
$ds^2_{(5)}(\pm r_0) = 0$ hypersurfaces at the
$r = \pm r_0$. In some sense these hypersurfaces are
like to the event horizon and in Ref.\cite{bron} such
hypersurfaces are named as a $D$-holes. On these
hypersurfaces we should join \cite{dzh7}:
\begin{itemize}
\item
the flux of the 4D electric field (defined by the Maxwell
equations) with the flux of the 5D electric field defined
by $R_{5t} = 0$ Kaluza-Klein equation.
\item
the area of the Reissner-Nordstr\"om event horizon with
the area of the $ds^2_{(5)}(\pm r_0) = 0$ hypersurface.
\end{itemize}
It is necessary to note that both solutions (Reissner-Nordstr\"om
black hole and 5D throat) have only two integration
constants\footnote{in fact, for the Reissner-Nordstr\"om
black hole this leads to the ``no hair'' theorem.} and
on the event horizon takes place an algebraic relation
between these 4D and 5D integration constants. Another
explanation of the fact that we use only two joining condition
is the following (see Ref.\cite{dzh99yb} for the more detailed
explanations): in some sense on the event horizon holds
a ``holography principle''. This means that in the presence
of the event horizon the 4D and 5D Einstein equations lead
to a reduction of the amount of initial data. For example
the Einstein - Maxwell equations for the Reissner-Nordstr\"om
metric
\begin{eqnarray}
ds^2 & = & \Delta dt^2 - \frac{dr^2}{\Delta} - r^2
\left (
d\theta ^2 + \sin^2 d\varphi ^2
 \right ) ,
\label{dzh-eq2-1}\\
A_\mu & = & \left (
\omega ,0,0,0)
 \right )
\label{dzh-eq2-2}
\end{eqnarray}
(where $A_\mu$ is the electromagnetic potential, $\kappa$
is the gravitational constant) can be written as
\begin{eqnarray}
-\frac{\Delta '}{r} + \frac{1 - \Delta}{r^2} & = &
\frac{\kappa}{2} {\omega '} ^2  ,
\label{dzh-eq2-3} \\
\omega ' & = & \frac{q}{r^2}     .
\label{dzh-eq2-4}
\end{eqnarray}
For the Reissner - Nordstr\"om black hole the event horizon
is defined by the condition $\Delta (r_g) = 0$, where $r_g$
is the radius of the event horizon.
Hence in this case we see that on the event horizon
\begin{equation}
\Delta '_g = \frac{1}{r_g} - \frac{\kappa}{2}
r_g {\omega '_g}^2  ,
\label{dzh-eq2-5}
\end{equation}
here (g) means that the corresponding value is taken
on the event horizon. Thus, Eq. (\ref{dzh-eq2-3}), which is the
Einstein equation, is a first-order differential equations
in the whole spacetime $(r \ge r_g)$. The condition (\ref{dzh-eq2-5})
tells us that the derivative of the metric on the event horizon
is expressed through the metric value on the event horizon.
This is the same what we said above: the reduction of the
amount of initial data takes place by such a way that we have
only two integration constants (mass $m$ and charge $e$ for the
Reissner-Nordstr\"om solution and $q$ and $r_0$ for the 5D
throat).
\par
The 5D throat has an interesting property \cite{vdschmidt2}.
We see that the signs of the $\eta_{55}$ and $\eta_{00}$ are not
defined. We remark that this 5D metric is located behind the event
horizon therefore the 4D observer is not able to determine the
signs of the $\eta_{55}$ and $\eta_{00}$. Moreover this 5D metric
can fluctuate between these two possibilities. Hence the external
4D observer is forced to describe such composite WH by means of
something like spinor.
\par
\begin{figure}
\centerline{ \framebox{
\psfig{figure=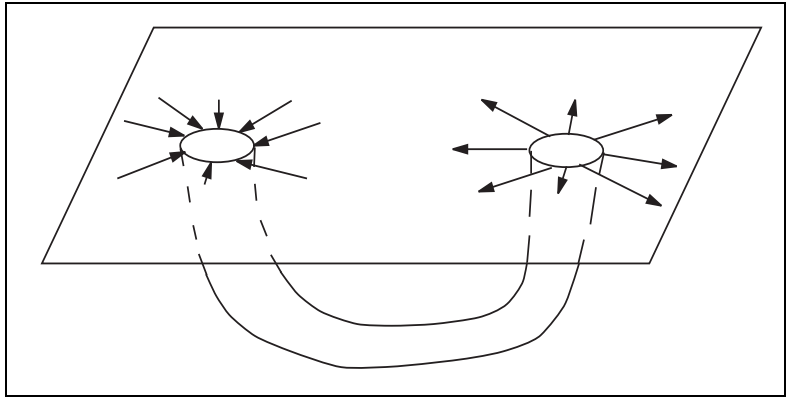,height=5cm,width=10cm}}}
\caption{The left mouth of the quantum WH entraps the force lines
of the electric field and looks as (-) electric charge.
The force lines outcome from the right mouth of WH which one
looks as (+) charge.}
\label{dzh-fig2a}
\end{figure}

Another interesting characteristic property of this solution
is that we have the flux of electric field through the throat,
{\it i.e.} each mouth can entrap the electric force lines
and this leads that
this mouth is like to electric charge for the external 4D
observer, see Fig.\ref{dzh-fig2a}.
We can neglect the cross section of the throat and in
this case each mouth is point-like and we can try to describe
these mouthes with help of some effective field. Taking into
account the spinor-like properties of quantum handles, we assume
that
\textit{\textbf{spacetime foam can be described with help
of an effective spinor field.}}

\section{Approximate model of the spacetime foam}

\begin{figure}
\centerline{ \framebox{
\epsfig{figure=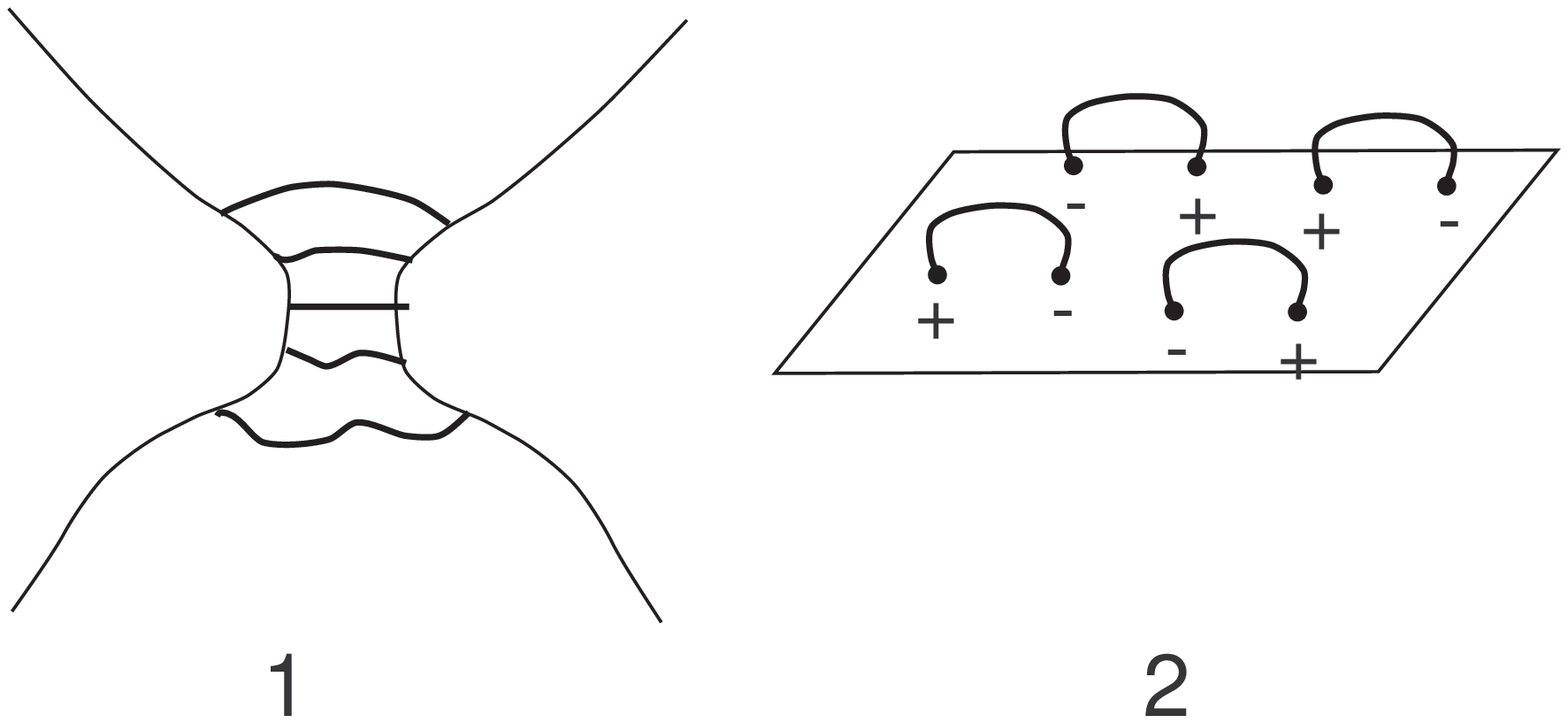,height=5cm,width=10cm}}}
\caption{At the left hand of the figure quantum handles
connect two spaces. At the right hand the mouthes of
quantum handles are separated in distance of the order
$l_{Pl}$.}
\label{dzh-fig3}
\end{figure}

The physical meaning of the spinor field depends on the method of
attaching the quantum handles to the external space, see
Fig.(\ref{dzh-fig3}).

\subsection{Quantum wormholes with separated mouthes}

In this case $|\psi|^2$ is a density of
the mouthes in the external space and $e|\psi|^2$ is a density of
the electric charge \cite{dzh00a}.
\par
Following this way we write differential equations for the
gravitational + electromagnetic fields in the presence of the
spacetime foam $(\psi)$ as follows
\begin{eqnarray}
R_{\mu\nu} -\frac{1}{2} g_{\mu\nu} R& = & T_{\mu\nu} ,
\label{dzh-eq3}\\
\left ( i \gamma^\mu \partial _\mu + e A_\mu -
\frac{i}{4}\omega _{\bar a\bar b\mu}\gamma^\mu \gamma^{[\bar
a}\gamma^{\bar b]} - m \right )\psi & = & 0 ,
\label{dzh-eq4}\\
D_\nu F^{\mu\nu} = 4\pi e \left ( {\bar\psi}\gamma^\mu  \psi \right
), &&
\label{dzh-eq5}
\end{eqnarray}
For our model we use the following ansatz: the spherically
symmetric metric
\begin{equation}
ds^2 = e^{2\nu(r)}\Delta (r) dt^2 - \frac{dr^2}{\Delta (r)} - r^2
\left ( d\theta ^2 + \sin ^2d\varphi ^2 \right ) ,
\label{dzh-eq6}
\end{equation}
the electromagnetic potential
\begin{equation}
A_\mu = \left ( -\phi,0,0,0 \right ) ,
\label{dzh-eq7}
\end{equation}
and the spinor field
\begin{equation}
\tilde \psi = e^{-i\omega t}\frac{e^{-\nu /2}}{r\Delta ^{1/4}}
\left ( f,0,ig\cos\theta ,ig\sin\theta e^{i\varphi} \right ) .
\label{dzh-eq8}
\end{equation}
The following is \textit{very important} for us: the ansatz
(\ref{dzh-eq8}) for the spinor field $\psi$ has the $T_{t\varphi}$
component of the energy-momentum tensor and the $J^\varphi = 4\pi
e (\bar\psi \gamma^\varphi \psi)$ component of the current. Let we
remind that $\psi$ determines the stochastical gas of the virtual WH's
which can not have a preferred direction in the spacetime. This
means that after substitution expression (\ref{dzh-eq6})-(\ref{dzh-eq8})
into field equations they should be averaged by the spin direction
of the ansatz (\ref{dzh-eq8}). After this averaging we have
$T_{t\varphi} = 0$ and $J^\varphi = 0$ and we have the following
equations system describing our spherically symmetric spacetime
\begin{eqnarray}
f' \sqrt{\Delta} & = & \frac{f}{r} - g
\left (
  \left (
  \omega - e\phi
  \right )\frac{e^{-\nu}}{\sqrt\Delta} + m
\right ) ,
\label{dzh-eq9}\\
g' \sqrt{\Delta} & = & f
\left (
  \left (
  \omega - e\phi
  \right )\frac{e^{-\nu}}{\sqrt\Delta} - m
\right ) -
\frac{g}{r} ,
\label{dzh-eq10}\\
r\Delta ' & = & 1 - \Delta -
\kappa \frac{e^{-2\nu}}{\Delta}
\left (\omega - e\phi \right )
\left (f^2 + g^2 \right ) - r^2e^{-2\nu}{\phi'}^2 ,
\label{dzh-eq11}\\
r\Delta\nu ' & = & \kappa\frac{e^{-2\nu}}{\Delta}
\left (\omega - e\phi \right )
\left (f^2 + g^2 \right ) -
\kappa\frac{e^{-\nu}}{r\sqrt\Delta}fg -
\nonumber \\
&& \frac{\kappa}{2}m\frac{e^{-\nu}}{\sqrt\Delta}\left (f^2 - g^2\right ) ,
\label{dzh-eq12}\\
r^2\Delta \phi '' & = & - 8\pi e
\left (f^2 + g^2 \right ) -
\left (
2r\Delta - r^2\Delta \nu '
 \right )\phi ' ,
\label{dzh-eq13}
\end{eqnarray}
where $\kappa$ is some constant.
This equations system was investigated in \cite{finster99} and
result is the following. A particle-like solution exists which
has the following expansions near $r = 0$
\begin{eqnarray}
f(r) & = & f_1r + {\cal O}(r^2),
\quad
g(r) = {\cal O}(r^2) ,
\label{dzh-eq14}\\
\Delta (r) & = & 1 + {\cal O}(r^2),
\quad
\nu(r) = {\cal O}(r^2) ,
\quad
\phi (r) = {\cal O}(r^2)
\label{dzh-eq15}
\end{eqnarray}
and the following asymptotical behaviour
\begin{eqnarray}
\Delta (r) & \approx & 1 - \frac{2m_\infty}{r} +
\frac{(2e_\infty)^2}{r^2}, \quad \nu (r) \approx const ,
\label{dzh-eq16}\\
\phi (r) & \approx & \frac{2e_\infty}{r} ,
\label{dzh-eq17}\\
f & \approx & f_0e^{-\alpha r}, \quad g \approx
g_0e^{-\alpha r},
\nonumber \\
\frac{f_0}{g_0} & = & \sqrt{\frac{m_\infty +
\omega}{m_\infty - \omega}}, \quad \alpha ^2 = m_\infty^2 - \omega
^2 ,
\label{dzh-eq18}
\end{eqnarray}
where $m_\infty$ is the mass for the observer at infinity and
$2e_\infty$ is the charge of this solution.
\par
The solution exists for both cases $(|e_\infty|/m_\infty)>1$ and
$(|e_\infty|/m_\infty)<1$ but for us is essential the first case with
$(|e_\infty|/m_\infty)>1$. In this case the classical Einstein-Maxwell theory
leads to the ``naked'' singularity. The presence of the spacetime
foam drastically changes this result:
\textbf{\textit{the appearance of
the virtual wormholes can prevent the formation
of the ``naked'' singularuty in the Reissner-Nordstr\"om solution
with $|e|/m > 1$}}.
\par
\begin{figure}
\centerline{ \framebox{
\psfig{figure=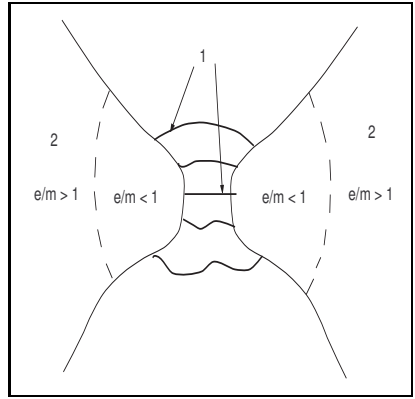,height=5cm,width=5cm}}}
\caption{{\bf 1} are the quantum (virtual) WHs, {\bf 2} are two
solutions with $|e_\infty|/m_\infty>1$. Such object can be named
as \textbf{\textit{the wormhole with quantum throat}.}}
\label{dzh-fig4}
\end{figure}

Our interpretation of this solution is presented on the
Fig.(\ref{dzh-fig4}).

\subsection{Quantum wormholes with non-separated mouthes}

\begin{figure}
\centerline{ \framebox{
\psfig{figure=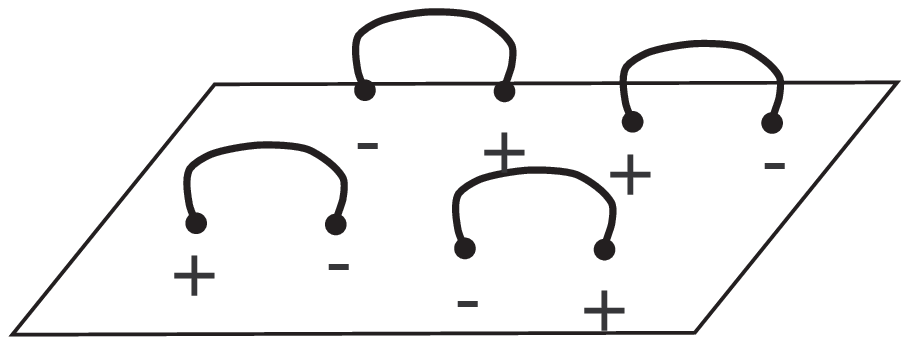,height=5cm,width=10cm}}}
\caption{The distance between mouthes of the quantum handle
is of order $l_{Pl}$.}
\label{dzh-fig5}
\end{figure}

The second possibitiy \cite{dzh00b} is presented on the
Fig.(\ref{dzh-fig5}).
\par
We will consider the 5D Kaluza-Klein theory + torsion +
spinor field. The Lagrangian in this case is
\begin{eqnarray}
{\cal L} & = & \sqrt{-G}
\left \{
-\frac{1}{2k}
  \left (
  R^{(5)} - S_{ABC}S^{ABC}
  \right ) +
\right.
\nonumber \\
&&
\left.
 \frac{\hbar c}{2}
  \left [
  i\bar\psi \left (\gamma^C
        \nabla _C - \frac{mc}{i\hbar}
        \right )\psi  +
        h.c.
  \right ]
\right\}
\label{dzh-eq19}
\end{eqnarray}
where $\nabla_C = \partial_C - \frac{1}{4}(\omega _{\bar A \bar B C}
+ S_{\bar A \bar B C})\gamma^{[\bar A}\gamma^{\bar B]}$
is the covariant derivative,
$G$ is the determinant of the 5D metric,
$R^{(5)}$ is the 5D scalar curvature, $S_{ABC}$ is the
antisymmetrical torsion tensor, $A,B,C$ are the 5D world indexes,
$\bar A, \bar B, \bar C$ are the 5-bein indexes,
$\gamma ^B = h^B_{\bar A}\gamma^{\bar A}$,
$h^B_{\bar A}$ is the 5-bein,
$\gamma^{\bar A}$ are the 5D $\gamma$ matrixes with usual definitions
$\gamma^{\bar A}\gamma^{\bar B} + \gamma^{\bar B}\gamma^{\bar A} =
2\eta^{\bar A\bar B}$,
$\eta^{\bar A\bar B} = (+,-,-,-,-)$ is the signature of the
5D metric, $\psi$ is the spinor field which effectively and
approximately describes the spacetime foam, $[]$ means the
antisymmetrization, $\hbar$, $c$ and $m$ are the usual constants.
After dimensional reduction
we have
\begin{eqnarray}
{\cal L} = \sqrt{-g}
\left \{
-\frac{1}{2k}
        \left (
        R + \frac{1}{4}F_{\alpha\beta}F^{\alpha\beta}
        \right ) +
\right . &&
\nonumber \\
\left.
\frac{\hbar c}{2}
        \left [
        i\bar \psi
                \left (
                \gamma^\mu \tilde\nabla_\mu -
                \frac{1}{8}F_{\bar\alpha\bar\beta}\gamma^{\bar 5}
                \gamma^{[\bar\alpha}\gamma^{\bar\beta ]} -
                \right .\right . \right . &&
\nonumber \\
                \left .\left . \left .
                \frac{1}{4}l^2_{Pl}
                \left (\gamma^{[\bar A}\gamma^{\bar B}\gamma^{\bar C]} \right )
                \left (i\bar\psi \gamma_{[\bar A}\gamma_{\bar B}\gamma_{\bar C]}
                \psi \right ) - \frac{mc}{i\hbar}
                \right )\psi  + h.c.
        \right ]
\right \} &&
\label{dzh-eq20}\\
S^{\bar A\bar B\bar C} =
2l^2_{Pl}
\left (
i\bar\psi \gamma^{[\bar A}\gamma^{\bar B}\gamma^{\bar C]}\psi
\right ) &&
\label{dzh-eq21}
\end{eqnarray}
where $g$ is the determinant of the 4D metric,
$\tilde \nabla _\mu = \partial _\mu -
\frac{1}{4} \omega _{\bar a\bar b \mu}
\gamma^{[\bar a} \gamma^{\bar b]}$ is the 4D covariant derivative
of the spinor field without torsion,
$R$ is the 4D scalar curvature,
$F_{\alpha\beta} = \partial_\alpha A_\beta - \partial_\beta A_\alpha$
is the Maxwell tensor,
$A_\mu = h^{\bar 5}_\mu$ is the electromagnetic potential,
$\alpha ,\beta ,\mu$ are the 4D world indexes,
$\bar\alpha ,\bar\beta, \bar\mu$ are the 4D vier-bein indexes,
$h^{\bar \mu}_\nu$ is the vier-bein,
$\gamma^{\bar \mu}$ are the 4D $\gamma$ matrixes with usual definitions
$\gamma^{\bar\mu}\gamma^{\bar\nu} + \gamma^{\bar\nu}\gamma^{\bar\mu} =
2\eta^{\bar\mu\bar\nu}$,
$\eta^{\bar\mu\bar\nu} = (+,-,-,-)$ is the signature of the
4D metric. Varying with respect to $g_{\mu\nu}$, $\bar\psi$
and $A_\mu$ leads to the following equations
\begin{eqnarray}
R_{\mu\nu} - \frac{1}{2}g_{\mu\nu}R =
\frac{1}{2}
\left (
-F_{\mu\alpha}F_\nu^\alpha + \frac{1}{4}g_{\mu\nu}
F_{\alpha\beta} F^{\alpha\beta}
 \right ) & + &
\nonumber \\
4l^2_{Pl}
\left [
        \left (
        i\bar\psi\gamma_\mu \tilde\nabla_\nu \psi +
        i\bar\psi\gamma_\nu \tilde\nabla_\mu \psi
         \right ) + h.c.
 \right ] & - &
\nonumber \\
2l^2_{Pl}
\left [
F_{\mu\alpha}
\left (i\bar\psi \gamma^{\bar 5} \gamma_{[\nu}
\gamma^{\alpha ]} \psi \right ) +
F_{\nu\alpha}
\left (i\bar\psi \gamma^{\bar 5} \gamma_{[\mu}
\gamma^{\alpha ]} \psi \right )
 \right ] & - &
 \nonumber \\
2g_{\mu\nu} l^4_{Pl}
\left (
i\bar\psi \gamma^{[\bar A}\gamma^{\bar B}\gamma^{\bar C]}\psi
\right )
\left (
i\bar\psi \gamma_{[\bar A}\gamma_{\bar B}\gamma_{\bar C]}\psi
\right ) & , &
\label{dzh-eq22}\\
D_\nu H^{\mu\nu} = 0 ,
\,
H^{\mu\nu} =
F^{\mu\nu} + \tilde F^{\mu\nu} & , &
\nonumber \\
\tilde F^{\mu\nu} = 4l^2_{Pl}
\left (
i\bar\psi \gamma^{\bar 5}\gamma^{[\mu} \gamma^{\nu ]}\psi
 \right ) =
4l^2_{Pl}
E^{\mu\nu\alpha\beta}
\left (
i\bar\psi \gamma_{[\alpha} \gamma_{\beta ]}\psi
 \right ) & , &
\label{dzh-eq23}\\
  i\gamma^\mu \tilde \nabla _\mu \psi-
  \frac{1}{8} F_{\bar \alpha \bar \beta}
  \left (
     i\gamma^{\bar 5}\gamma^{[\bar \alpha} \gamma^{\bar \beta ]} \psi
  \right ) & - &
\nonumber \\
  \frac{1}{2} l^2_{Pl}
  \left (
     i\gamma^{[\bar A}\gamma^{\bar B}\gamma^{\bar C]} \psi
  \right )
  \left (
     i\bar\psi \gamma_{[\bar A}\gamma_{\bar B}\gamma_{\bar C]}\psi
  \right ) & = & 0 ,
\label{dzh-eq24}
\end{eqnarray}
where $\omega _{\bar a\bar b \mu}$ is the 4D Ricci coefficients without
torsion, $E^{\mu\nu\alpha\beta}$ is the 4D absolutely antisymmetric
tensor. The most interesting for us is the Maxwell equation
(\ref{dzh-eq23}) which permits us to discuss the physical meaning
of the spinor field.
We would like to show that this equation in the given form
is similar to the electrodynamic in the continuous media.
Let we remind that for the electrodynamic in the continuous media
two tensors $\bar F^{\mu\nu}$ and $\bar H^{\mu\nu}$ are introduced
\cite{landau} for which we have the following equations system
(in the Minkowski spacetime)
\begin{eqnarray}
\bar F_{\alpha\beta ,\gamma} +
\bar F_{\gamma\alpha ,\beta} +
\bar F_{\beta\gamma ,\alpha} & = & 0 ,
\label{dzh-eq26}\\
\bar H^{\alpha\beta}_{,\beta} & = & 0
\label{dzh-eq27}
\end{eqnarray}
and the following relations between these tensors
\begin{eqnarray}
\bar H_{\alpha\beta} u^\beta & = &
\varepsilon \bar F_{\alpha\beta}u^\beta ,
\label{dzh-eq28}\\
\bar F_{\alpha\beta} u_\gamma +
\bar F_{\gamma\alpha} u_\beta +
\bar F_{\beta\gamma} u_\alpha & = &
\mu
\left (
\bar H_{\alpha\beta} u_\gamma +
\bar H_{\gamma\alpha} u_\beta +
\bar H_{\beta\gamma} u_\alpha
 \right )
\label{dzh-eq29}
\end{eqnarray}
where $\varepsilon$ and $\mu$ are the dielectric
and magnetic permeability respectively, $u^\alpha$ is the
4-vector of the matter. For the
rest media and in the 3D designation we have
\begin{eqnarray}
\varepsilon \bar E_i & = & \bar E_i + 4\pi \bar P_i = \bar D_i ,
\quad \mbox{where} \quad
\bar E_i = \bar F_{0i} ,
\quad
\bar D_{0i} = \bar H_{0i} ,
\label{dzh-eq30}\\
\mu \bar H_i & = &
\bar H_i + 4\pi \bar M_i = \bar B_i ,
\quad \mbox{where} \quad
\bar B_i = \epsilon_{ijk}\bar F^{jk} ,
\quad
\bar H_i = \epsilon_{ijk}\bar H^{jk} ,
\label{dzh-eq31}
\end{eqnarray}
where $P_i$ is the dielectric polarization and $M_i$ is the
magnetization vectors, $\epsilon_{ijk}$ is the 3D absolutely
antisymmetric tensor. Comparing with the (\ref{dzh-eq23}) Maxwell
equation for the spacetime foam in the 3D form
\begin{eqnarray}
E_i + \tilde E_i & = & D_i
\quad \mbox{where }\quad
E_i = F_{0i}, \quad \tilde E_i = \tilde F_{0i} ,
\quad D_i = H_{0i}
\label{dzh-eq32}\\
B_i + \tilde B_i & = & H_i
\quad \mbox{where} \;
B_i = \epsilon_{ijk} F^{jk},
\;
\tilde B_i = \epsilon_{ijk} \tilde F^{jk},
\;
H_i = \epsilon_{ijk} H^{jk}
\label{dzh-eq33}
\end{eqnarray}
we see that the following notations can be introduced.
\begin{equation}
\tilde E_i = 4l^2_{Pl} \epsilon_{ijk}
\left (
i\bar \psi \gamma^{[j} \gamma^{k]} \psi
 \right )
\label{dzh-eq34}
\end{equation}
is the polarization vector of the spacetime foam and
\begin{equation}
\tilde B_i = -4l^2_{Pl} \epsilon_{ijk}
\left (
i\bar \psi \gamma ^{\bar 5} \gamma^{[j} \gamma^{k]} \psi
 \right )
\label{dzh-eq35}
\end{equation}
is the magnetization vector of the spacetime foam.
\par
\begin{figure}
\centerline{ \framebox{
\epsfig{figure=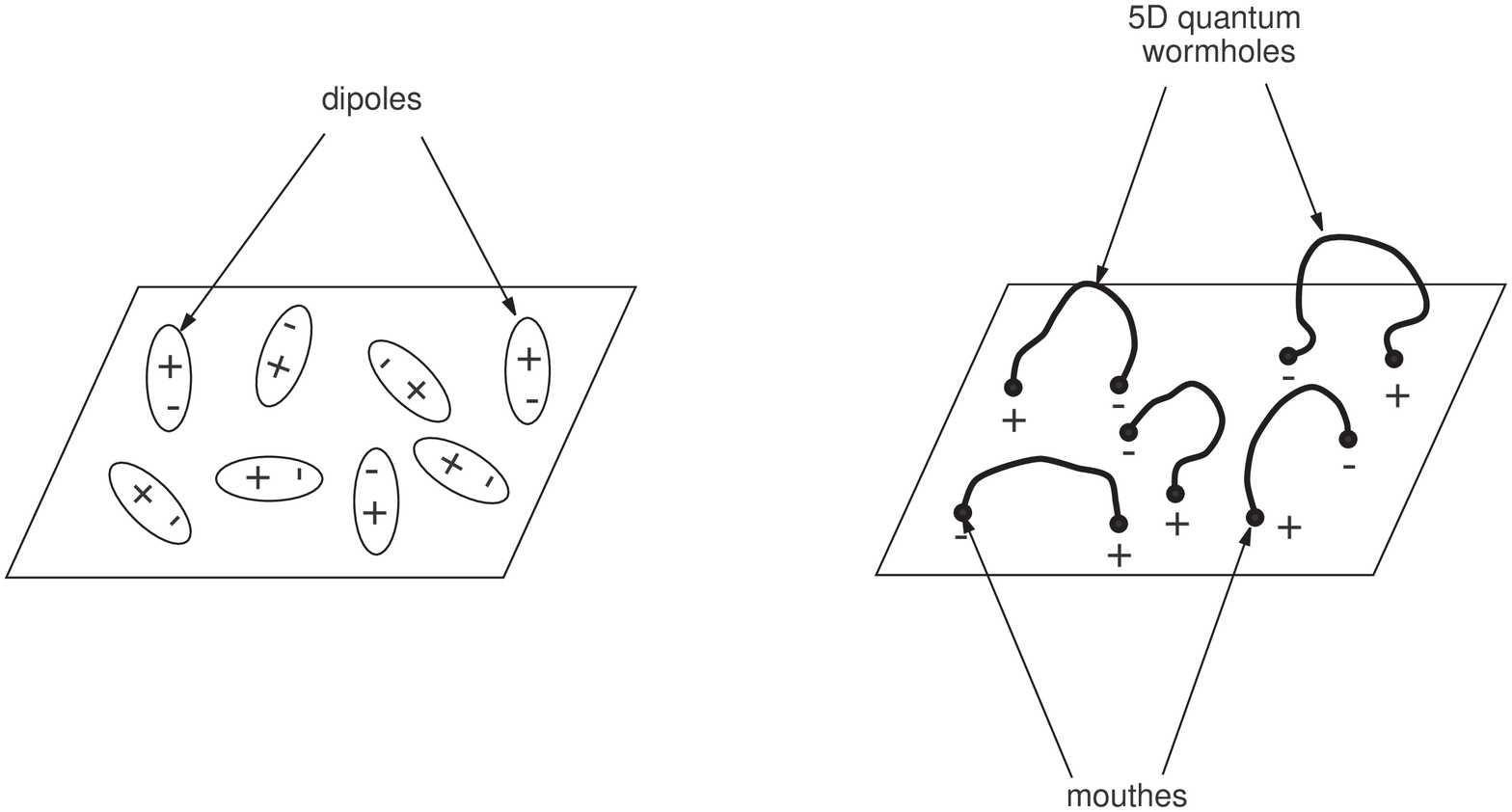,height=5cm,width=10cm}}}
\caption{For the 4D observer each mouth looks as a moving electric
charge. This allows us in some approximation imagine the
spacetime foam as a continuous media with a polarization.}
\label{dzh-fig6}
\end{figure}

The physical reason for this is evidently: each quantum WH is like to a
moving dipole (see Fig.(\ref{dzh-fig6}) which produces microscopical
electric and magnetic fields.

\section{Supergravity as a possible model of the spacetime foam}

From the above-mentioned arguments we see that the most important
for such kind models of the spacetime foam is the presence of the
nonminimal interaction term (in Lagrangian) between spinor and
electromagnetic fields. Let we note that the N=2 supergravity
\cite{ferrara76}
which contains the vier-bein $e^a_\mu$, Majorana Rarita-Schwinger
field $\psi_\mu$, photon $A_\mu$ and a second Majorana
spin-$\frac{3}{2}$ field $\varphi_\mu$ has the following term in
Lagrangian
\begin{eqnarray}
{\cal L}_{se} & = & \frac{\kappa}{\sqrt 2} \bar \psi _\mu \left ( e
F^{\mu \nu} + \frac{1}{2} \gamma_5 \tilde F^{\mu \nu} \right )
\varphi_\nu + \cdots ,
\nonumber \\
\quad \tilde F_{\mu \nu} & = &
e_{\mu\nu\alpha\beta} F^{\alpha \beta}
\label{dzh-eq36}
\end{eqnarray}
The term like this usually occur in supergravities which have some
gauge multiplet of supergravity and some matter multiplet. Taking
into account the previous reasonings we can suppose that
\textbf{
\textit{supergravity theories can be considered as approximate
models of the spacetime foam.}}

\section{Conclusions}

Thus, here we have proposed the approximate model for the
description of the spacetime foam. This model is based on the
assumption that the whole spacetime is 5 dimensional
but $G_{55}$ is the dynamical variable only in the quantum
topological handles (wormholes).
In this case 5D gravity has the solution which we have used as a
model of the single quantum wormhole. The properties of
this solution is such that we can assume that the
quantum topological handles (wormholes) can be
approximately described by some effective spinor field.
\par
The topological handles of the spacetime foam either can be
attached to one space or connect two different spaces. In the
first case we have something like to strings between two
$D$-branes (or wormhole with the quantum throat) and such
object can demonstrate a model of preventing the formation
the naked singularity with relation $e > m$. In the second
case the spacetime foam looks as a dielectric with quantum
handles as dipoles.
\par
Such model leads to the very interesting experimental consequences.
We see that the spacetime foam has 5D structure and it connected
with the electric field. This observation allows us to presuppose
that the very strong electric field can open a door into 5
dimension! The question is: as is great should be this field ? The
electric field $E_i$ in the CGSE units and $e_i$ in the ``geometrized''
units can be connected by formula
\begin{eqnarray}
e_i & = & \frac{G^{1/2}}{c^2}E_i =
\left (
2.874 \times 10^{-25} \; cm^{-1}/gauss
 \right )E_i ,
\label{dzh-eq37}\\
\left [ e_i \right ] & = & cm^{-1} ,
\quad
\left [E_i\right ] = V/cm
\label{dzh-eq38}
\end{eqnarray}
As we see the value of $e_i$ is defined by some characteristic
length $l_0$. It is possible that $l_0$ is a length of the
$5^{th}$ dimension. If $l_0 = l_{Pl}$ then
$E_i \approx 10^{57} V/cm$ and this field strength is in the
Planck region, and is will beyond experimental capabilities
to create. But if $l_0$ has a different value it can lead to
much more realistic scenario for the experimental capability
to open door into $5^{th}$ dimension.
\par
Another interesting conclusion of this paper is that
supergravity theories having nonminimal interaction
between spinor and electromagnetic fileds
\textit\textbf{{can be considered as approximate and
effective models of the spacetime foam}}.

\section{Acknowledgment}

I would like to acknowledze the generosity of NATO
in its support for this workshop and ICTP (grant KR-154).

\end{article}

\end{document}